\documentstyle{mn}
\input{psfig}

\begin{document}


\def\ergsec{\hbox{erg s$^{-1}$}}
\def\ergcmsec{\hbox{erg cm$^{-2}$ s$^{-1}$}}
\def\countsec{\hbox{counts s$^{-1}$}}
\def\photcmsec{\hbox{photon cm${-2}$ s$^{-1}$}}
\def\degmark{^\circ}
\def \rsun {\ifmmode$R$_{\odot}\else R$_{\odot}$\fi}
\def \nh {N$_{\rm H}$}
\def\ax{$\alpha_{\rm x}$}
\def \hcm {\hbox {\ifmmode $ H atoms cm$^{-2}\else H atoms cm$^{-2}$\fi}}
\def\approxgt{\mathrel{\hbox{\rlap{\lower.55ex \hbox {$\sim$}}
        \kern-.3em \raise.4ex \hbox{$>$}}}}
\def\approxlt{\mathrel{\hbox{\rlap{\lower.55ex \hbox {$\sim$}}
        \kern-.3em \raise.4ex \hbox{$<$}}}}
\def\la{\mathrel{\hbox{\rlap{\hbox{\lower4pt\hbox{$\sim$}}}\hbox{$<$}}}}
\def\ga{\mathrel{\hbox{\rlap{\hbox{\lower4pt\hbox{$\sim$}}}\hbox{$>$}}}}
\def\kms{${\rm km~s^{-1}}$}
\def\PL{${\rm P_{1.4GHz}}$}      
\def\Lel{${\rm L_{e.l.}}$}
\def\uv{{\it uv\ }}
\def\FR{Fanaroff-Riley}
\newcommand {\rosat} {{ROSAT }}
\newcommand {\einstein} {{\it Einstein }}
\newcommand {\exosat} {{EXOSAT }}
\newcommand {\asca} {{\it ASCA }}
\newcommand {\sax} {{\it BeppoSAX }}
\newcommand {\ginga} {{\it GINGA }}
\newcommand {\ie} {{\it  i.e.}}
\newcommand {\cf} {{\it  cf }}
\newcommand {\eg} {{e.g.}}
\newcommand {\etal} {et~al. }
\newcommand {\Msun} {{ M$_{\odot}$}}
\newcommand {\degree} {$^{\circ}$}
\newcommand {\sqcm} {cm$^{2}$}
\newcommand {\cbcm} {cm$^{3}$}
\newcommand {\persqcm} {cm$^{-2}$}
\newcommand {\percbcm} {cm$^{-3}$}
\newcommand {\s} {s$^{-1}$}
\newcommand {\gps} {g s$^{-1}$}
\newcommand {\kmps} {km s$^{-1}$}
\newcommand {\lts} {{\it lt}-s}
\newcommand {\yr} {yr$^{-1}$}
\newcommand {\cps} {counts~s$^{-1}$}
\newcommand {\ergs} {erg~s$^{-1}$}
\newcommand {\ergcms} {erg cm$^{-2}$ s$^{-1}~$}
\newcommand {\chisq} {$\chi ^{2}$}
\newcommand {\rchisq} {$\chi_{\nu} ^{2}$}
\newcommand {\cpskeV} {counts s$ ^{-1}$ keV$ ^{-1}$ }
\newcommand{\magg}{\hphantom{$>$}}  
\newcommand{\dig}{\hphantom{0}}
\newcommand{\DXDYCZ}[3]{\left( \frac{ \partial #1 }{ \partial #2 }
                        \right)_{#3}}
\title[\sax Observations of 1-Jy BL Lacertae Objects. I.]
{\sax Observations of 1-Jy BL Lacertae Objects. I.}

\author[P. Padovani et al.]{Paolo Padovani$^{1,2,3,}$\thanks{Email: padovani@stsci.edu}, Luigi Costamante$^{4}$, Paolo Giommi$^{5}$, Gabriele Ghisellini$^6$, 
\newauthor  Andrea Comastri$^{7}$, Anna Wolter$^8$, Laura Maraschi$^8$, Gianpiero Tagliaferri$^6$, 
\newauthor C. Megan Urry$^1$
\\
$^1$ Space Telescope Science Institute, 3700 San 
Martin Drive, Baltimore MD. 21218, USA \\ 
$^2$ Affiliated to the Astrophysics Division, Space Science Department, 
European Space Agency \\
$^3$ On leave from Dipartimento di Fisica, II Universit\`a di Roma
``Tor Vergata'', Via della Ricerca Scientifica 1, I-00133 Roma, Italy \\
$^4$ Universit\`a degli Studi di Milano, Milano, Italy\\
$^5$ BeppoSAX Science Data Center, ASI, Via Corcolle 19, I-00131 Roma, Italy\\ 
$^6$ Osservatorio Astronomico di Brera, Via Bianchi 46, I-23807 Merate, Italy\\
$^7$ Osservatorio Astronomico di Bologna, Via Ranzani 1, I-40127 Bologna, Italy\\
$^8$ Osservatorio Astronomico di Brera, Via Brera 28, I-20121 Milano, Italy\\
}

\date{Accepted~~, Received~~}

\maketitle

\begin{abstract}
We present new \sax observations of seven BL Lacertae
objects selected from the 1 Jy sample plus one additional source. The
collected data cover the energy range $0.1 - 10$ keV (observer's frame),
reaching $\sim 50$ keV for one source (BL Lac). All sources
characterized by a peak in their multifrequency spectra at infrared/optical
energies (i.e., of the LBL type) display a relatively flat ($\alpha_{\rm x}
\sim 0.9$) X-ray spectrum, which we interpret as inverse Compton emission.
Four objects (2/3 of the LBL) show some evidence for a low-energy steepening
which is likely due to the synchrotron tail
merging into the inverse Compton component around $\sim 1 - 3$ keV. If this
were generally the case with LBL, it would explain why the $0.1 -2.4$ keV
ROSAT spectra of our sources are systematically steeper than the \sax ones
($\Delta \alpha_{\rm x} \sim 0.5$). The broad-band spectral energy
distributions fully confirm this picture and a synchrotron
inverse Compton model allows us to derive the physical parameters (intrinsic
power, magnetic field, etc.) of our sources. Combining our results with those
obtained by \sax on BL Lacs covering a wide range of synchrotron peak
frequency, $\nu_{\rm peak}$, we confirm and clarify the dependence of the
X-ray spectral index on $\nu_{\rm peak}$ originally found in ROSAT data.
\end{abstract} 


\section{Introduction}

BL Lacertae objects constitute one of the most extreme classes of active
galactic nuclei (AGN), distinguished by their high luminosity, rapid
variability, high ($> 3\%$) optical polarization, radio core-dominance,
apparent superluminal speeds, and almost complete lack of emission lines
(e.g., Kollgaard 1994; Urry \& Padovani 1995). The broad-band emission in
these objects, which extends from the radio to the gamma-ray band, appears
to be dominated by non-thermal processes from the heart of the AGN, undiluted
by the thermal emission present in other AGN. Therefore, BL Lacs represent the
ideal class to study to further our understanding of non-thermal emission 
in AGN.

Synchrotron emission combined with inverse Compton scattering is generally
thought to be the mechanism responsible for the production of radiation over
such a wide energy range (e.g., Ghisellini et al. 1998). The synchrotron peak
frequency, $\nu_{\rm peak}$, ranges across several orders of magnitude, going
from the far-infrared to the hard X-ray band (e.g., Sambruna et al. 1996;
Fossati et al. 1998). Sources at the extremes of this
wide distribution are referred to as low-energy peaked (LBL) and high-energy
peaked (HBL) BL Lacs, respectively (Giommi \& Padovani 1994; Padovani \& Giommi
1995). Radio-selected samples include mostly objects of the LBL type, while
X-ray selected samples are mostly made up of HBL. 

This scenario gives clear and strong predictions on the X-ray spectra for the
two classes. In the relatively narrow ROSAT band differences between the two
classes became apparent only when very large BL Lac samples ($\sim 50\%$ of
the then known objects) were considered (Padovani \& Giommi 1996; Lamer,
Brunner \& Staubert 1996). The \sax satellite (Boella \etal 1997a), with its
broad-band X-ray ($0.1-200$ keV) spectral capabilities, is particularly well
suited for a detailed analysis of the individual X-ray spectra of these 
sources.

In this paper we present \sax observations of eight BL Lacs, including six LBL
and two HBL. The sample is well defined (in particular, it is not a
compilation of known hard X-ray sources) being extracted, apart from one
source, from the radio-selected 1-Jy sample, for which a wealth of information
at many wavelengths is available.

In \S~2 we present our sample, \S~3 discusses the observations and the data
analysis, while \S~4 describes the results of our spectral fits to the \sax
data. \S~5 deals with the ROSAT data for our sources, while \S~6 presents the
spectral energy distributions and synchrotron-inverse Compton fits to the data,
and \S~7 deals with the dependence of the X-ray spectral index on synchrotron
peak frequency. Finally, \S~8 discusses our conclusions. Throughout this paper
spectral indices are written $S_{\nu} \propto \nu^{-\alpha}$.

\section{The Sample}
The 1 Jy sample of BL Lacs is presently the only sizeable, complete sample of
radio bright BL Lacs. It includes 34 objects with radio flux $> 1$ Jy at 5 GHz
(Stickel et al. 1991). All 1 Jy BL Lacs have been studied in detail in the
radio and optical bands; all objects have also soft X-ray data,
primarily from ROSAT.

We selected for \sax observations all 1 Jy BL Lacs with $0.1 - 10$ keV X-ray
flux larger than $2 \times 10^{-12}$ erg cm$^{-2}$ s$^{-1}$ (estimated from an
extrapolation of the single power-law fits derived for these objects from 
ROSAT data; Urry et al. 1996). This included twenty 1 Jy BL Lacs (or $\sim
60\%$ of the sample). Some of these sources have been included in other \sax
programs (e.g., MKN 501: Pian et al. 1998; S5 0716+714: Giommi et al. 1999). 
We present here the results
obtained for the 7 objects observed in Cycle 1 plus B2 0912+29, 
which was included in a program aimed at studying ``intermediate'' BL Lacs. 
All sources apart from PKS 2005$-$489 and B2 0912+29 
are LBL. The object list and basic characteristics are given 
in Table 1, which presents the source name, type, position, redshift, $R$ 
magnitude, 5 GHz radio flux, and Galactic \nh. 

\begin{table*}
{\bf Table 1. Sample Properties}
\begin{center}
\begin{tabular}{llrrllrl}
\hline
Name & Type & RA(J2000) & Dec(J2000)&~~~$z$ &R$_{\rm mag}^a$&F$_{\rm 5GHz}$&Galactic N$_{\rm H}$ \\
              &    &         &        &    &   &Jy &$10^{20}$ cm$^{-2}$ \\
\hline
PKS 0048$-$097&LBL & 00 50 41.3 & $-$09 29 06 &$>$0.2 & 16.5 & 2.0 & 3.85 \\
OJ 287        &LBL & 08 54 49.0 & $+$20 06 32 &\magg0.306 & 15.0 & 2.6 & 2.75$^{b}$ \\
B2 0912+29    &HBL & 09 15 52.3 & $+$29 33 21 &\magg  ...   & 15.8$^c$ & 0.2 & 2.11 \\   
PKS 1144$-$379 &LBL & 11 47 01.4 & $-$38 12 10 &\magg1.048 & 16.5 & 1.6 & 7.64 \\
PKS 1519$-$273 &LBL & 15 22 37.7 & $-$27 30 10 &$>$0.2 & 18.5 & 2.4 & 8.66 \\
4C 56.27      &LBL & 18 24 07.2 & $+$56 51 00 &\magg0.664 & 18.5 & 1.7 & 4.16 \\
PKS 2005$-$489 &HBL & 20 09 25.4 & $-$48 49 55 &\magg0.071 & 13.5 & 1.2 & 5.08 \\
BL Lac        &LBL & 22 02 43.2 & $+$42 16 40 &\magg0.069 & 14.0 & 4.8 & 20.15$^{b}$, total ~36$^{d}$ \\
\hline
\multicolumn{6}{l}{\footnotesize{$^a$ mean R magnitude from Heidt \& Wagner
(1996)}} \\
\multicolumn{6}{l}{\footnotesize{$^b$ from Elvis et al. 1989}} \\
\multicolumn{6}{l}{\footnotesize{$^c$ V magnitude from Tapia et al. (1976)}} \\
\multicolumn{6}{l}{\footnotesize{$^d$ including a molecular cloud along the 
line of sight (see Sambruna et al. 1999)}} \\
\end{tabular}
\end{center}
\end{table*}

\begin{table*}
{\bf Table 2. BeppoSAX Journal of observations}
\begin{tabular}{lrcrcrcc}
\hline
Name &LECS& LECS &MECS & MECS & PDS & PDS & Observing date\\
 &exp. (s)& count rate$^a$ (cts/s)&exp. (s)& count rate$^{a}$ (cts/s)  & 
 exp. (s)& count rate$^a$ (cts/s)& \\
\hline
PKS 0048$-$097& \dig4602&$0.020\pm0.006 $& \dig9810& $0.023\pm0.002 $ & ...  & ... & 1997 Dec 19 \\
OJ 287        & \dig5105&$0.018\pm0.006 $&    10707& $0.032\pm0.002 $ & 4511 & $0.128\pm0.094$ & 1997 Nov 24 \\
B2 0912+29    & \dig9195&$0.054\pm0.004 $&    23832& $0.048\pm0.002 $ & ...  & ... & 1997 Nov 14-15 \\
PKS 1144$-$379&    10649&$0.010\pm0.003 $&    22754& $0.019\pm0.001 $ & ...  & ... & 1997 Jan 10-11 \\
PKS 1519$-$273& \dig9266&$0.009\pm0.003 $&    26906& $0.009\pm0.001 $ & ...  & ... & 1998 Feb 1 \\
4C 56.27      & \dig4104&$0.014\pm0.007 $&    13382& $0.015\pm0.001 $ & ...  & ... & 1997 Oct 11 \\
PKS 2005$-$489& \dig... &  ...           & \dig9853& $1.371\pm0.012 $ & ...  & ... & 1996 Sep 29-30 \\
BL Lac        &    12260&$0.094\pm0.003 $&    13750& $0.150\pm0.003 $ & 8278 & $0.191\pm0.094$ & 1997 Nov 8 \\
\hline
\multicolumn{8}{l}{\footnotesize $^a$ net count rate full band} \\
\end{tabular}
\end{table*}

\section{Observations and Data Analysis}

A complete description of the \sax mission is given by Boella \etal (1997a). 
The relevant instruments for our observations are the coaligned Narrow Field
Instruments (NFI), which include one Low Energy Concentrator Spectrometer
(LECS; Parmar \etal 1997) sensitive in the 0.1 -- 10 keV band; three identical
Medium Energy Concentrator Spectrometers (MECS; Boella \etal 1997b), covering
the 1.5 -- 10 keV band; and the Phoswich Detector System (PDS; Frontera \etal
1997), coaligned with the LECS and the MECS. The PDS instrument is made up of
four units, and was operated in collimator rocking mode, with a pair of units
pointing at the source and the other pair pointing at the background, the two
pairs switching on and off source every 96 seconds. The net source spectra
have been obtained by subtracting the `off' to the `on' counts.  A journal of
the observations is given in Table 2. 

The data analysis was based on the linearized, cleaned event files obtained
from the \sax Science Data Center (SDC) on-line archive (Giommi \& Fiore 1997).
The data from the three MECS instruments were merged in one 
single event file by the SDC, based on sky coordinates.
The event file was then screened with a time filter given by SDC to
exclude those intervals related to events without attitude solution 
(i.e., conversion from detector to sky coordinates; 
 see Handbook for NFI Spectral Analysis, F. Fiore et al., 1999).
This was done to avoid an artificial decrease in the flux. 
As recommended by the SDC, the channels 1--10 and above 4 keV for the LECS and
0--36 and 220--256 for the MECS  were excluded from the spectral
analysis, due to residual calibration uncertainties.
Except for PKS 1519$-$273, where an extraction radius of 
4 arcmin was used for the LECS (see below), spectra and lightcurves 
have been extracted  using the standard extraction radii of
8 and 4 arcmin, for the LECS and MECS respectively. 

The spectral analysis was performed using the matrices and blank-sky
background files released in November 1998 by the SDC. For the LECS, we were
careful to choose the blank-sky file extracted in the same coordinate frame as
the source file. This is necessary to avoid an error in the background
subtraction that arises when using source and background files extracted in
coordinate frames of different pixel size (in this case, raw and detector or
sky coordinates). Raw pixels, in fact, have a size of 14 arcsec, while
detector and sky pixels have a size of 8 arcsec. Therefore, an equal
extraction region of 8 arcmin, for example, is obtained with a different
number of pixels. In this situation, however, the spectral files extracted in
the two cases have different values of the keyword used in the
analysis software to rescale the background to the source extraction area. The
two files, in short, appear to have been extracted with different areas (the
difference is $\sim35\%$), and the background in such case would be wrongly
rescaled accordingly.

Because
of the importance of the band below 1 keV to assess the presence of extra
absorption or soft excess (indicative of a double power-law spectrum), we have
also checked the LECS data for differences in the cosmic background between local
and blank-sky field observations, comparing spectra extracted from the 
same areas on the detector (namely, two circular regions outside the
$10^{\prime}$ radius central region, located at the opposite corners with
respect to the two on-board radioactive calibration sources).  
No relevant differences were found, except in the BL Lac and PKS
1519$-$273 observations.

For BL Lac, the blank-sky background up to 0.5--0.6 keV is higher than 
that during our
observation, causing a low level of counts in the source
spectrum. 
The signal-to-noise ratio (S/N) at low energy, however, is very low, due to 
the high galactic 
absorption ($\sim2-3.6\times 10^{21}$ cm$^{-2}$, see Tab. 1), and 
no detection is expected below $0.3-0.4$ keV.
In a conservative approach, we have therefore limited the spectral
analysis to energies above 0.5 keV. 
However, we also checked the results down to 0.1 keV 
using a different background file, obtained multiplying the blank-sky
field at the source position by the ratio of the local to blank field
backgrounds extracted in the two areas far from the source. This should give
an estimate of the local background at the source position. No significant
differences or trends between the two cases were found.

For PKS 1519$-$273, the local background presents an anomalously high flux
between $\sim$ 0.7 and 1 keV. This feature is also present in the
backgrounds extracted from the two circular regions independently, so
it is unlikely that its cause could be a very faint serendipitous source. This
problem becomes more evident with larger extraction radii, due to the very low
flux of the source. To minimize this effect and increase the S/N,
an extraction radius of 4 arcmin was used. Minor differences were found for
PKS 1144$-$379: the local background was uniformly higher by $\sim20\%$, so the
blank field background was rescaled accordingly, as suggested by the SDC 
(see Handbook for NFI spectral Analysis).

\subsection{Time analysis}

\begin{figure}
\centerline{\psfig{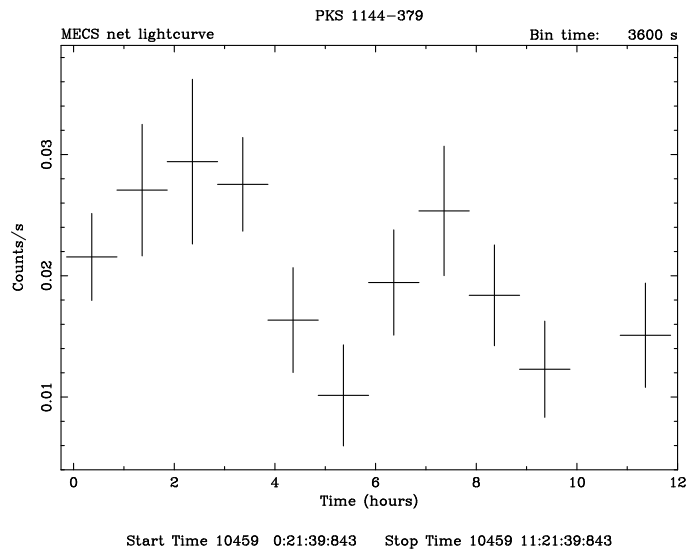}}
\caption{The net source lightcurve of the three MECS units for PKS 1144$-$379.
The apparent variability is significant at the $97\%$ level.}
\end{figure}

Using the software package XRONOS we looked for time variability in every
observation, binning the data in intervals from 500 to 4000 s, with null
results except for PKS 1144$-$379. In this case, there might be an indication
of variability, as the lightcurves present a ``wave-like'' shape, in particular
on time-scales $\sim 5$ hours. The resulting \chisq~value is consistent with no
variability at the $\sim 3$\% level so this result is not compelling, but the
same pattern is present in every single MECS detector, within the
uncertainties. This level remains roughly constant across different binnings,
given sufficient statistics (i.e., with time bins $\ga 1200$ s). Also
considering the whole sample, the significance is ``borderline": a spurious
variability of this level is expected, on average, every $\sim 33$
observations, in the hypothesis of a parent population of constant sources. In
our case this translates to 0.24 times every 8 sources. The LECS light curve 
is much less sampled and consistent with no variability
at the $\sim 17$\% level, although similar in shape to the MECS one. An inspection 
of the local background lightcurve
showed no significant variation. Fig. 1 shows the net source lightcurve of
the three MECS units merged together. If the variations are real, the source
varied up to a factor $\sim 3$ in 4 hours. 

\begin{figure}
\centerline{\psfig{figure=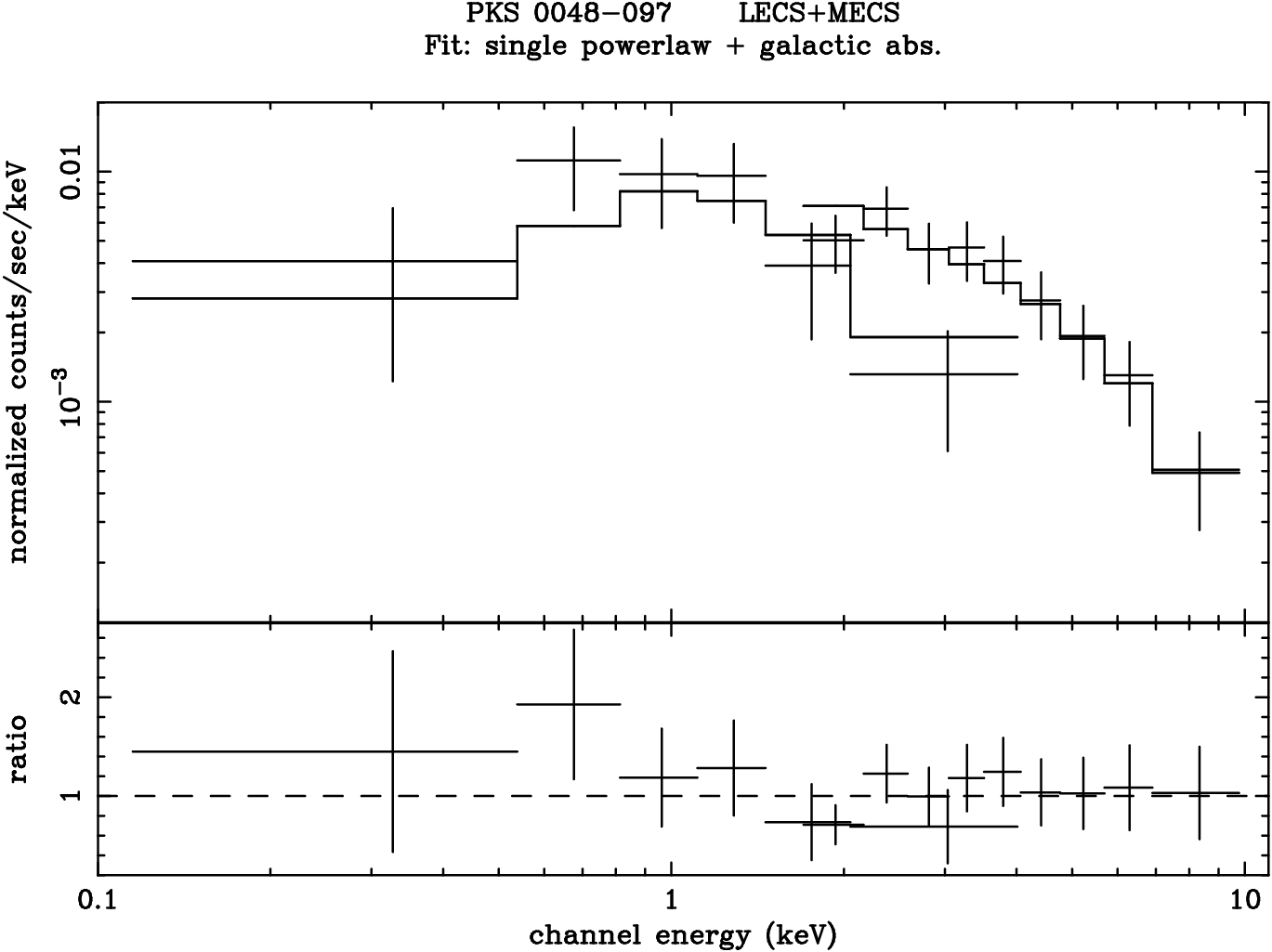,width=9cm,angle=-90}}
\medskip
\centerline{\psfig{figure=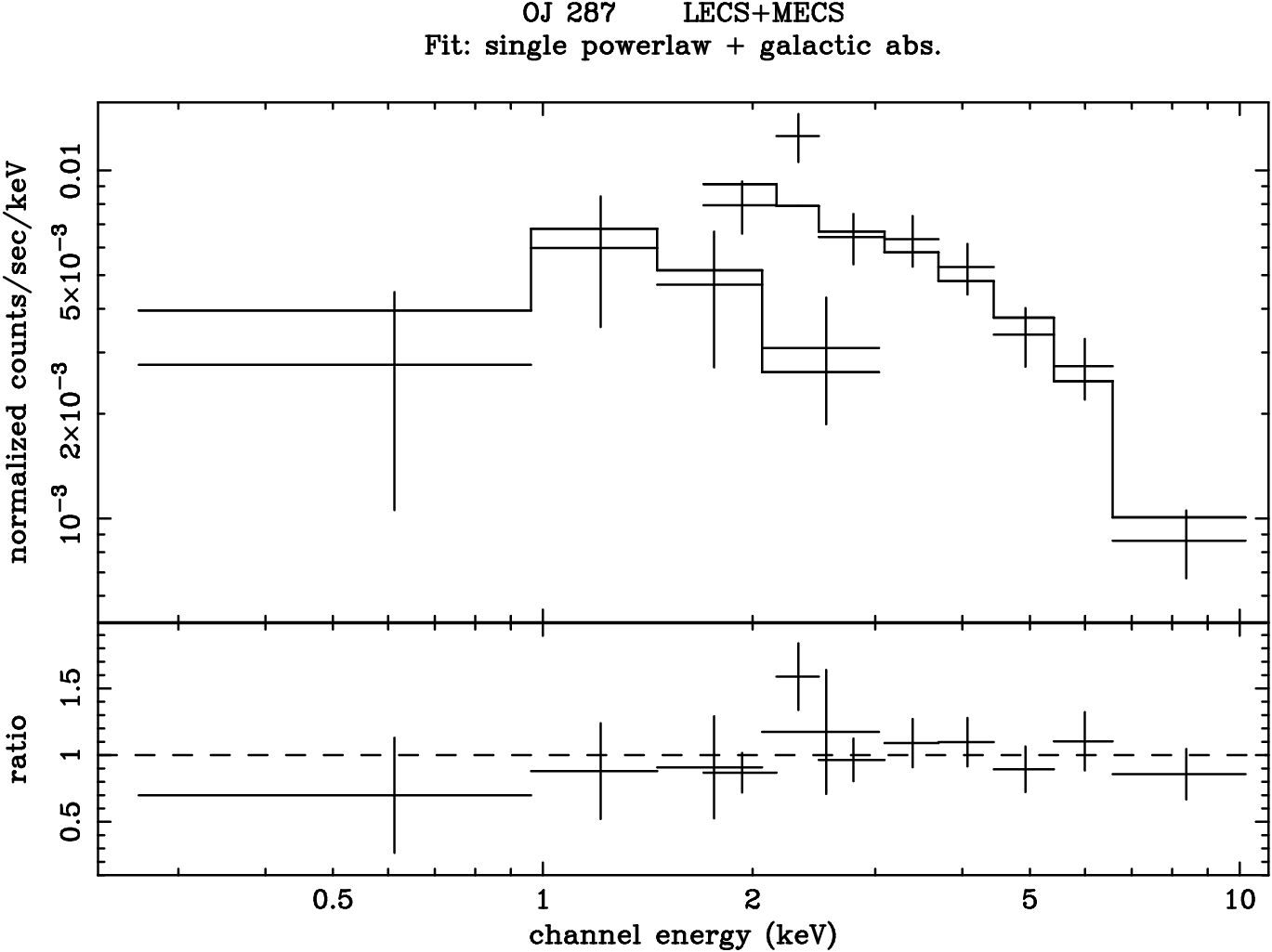,width=9cm,angle=-90}}
\medskip
\centerline{\psfig{figure=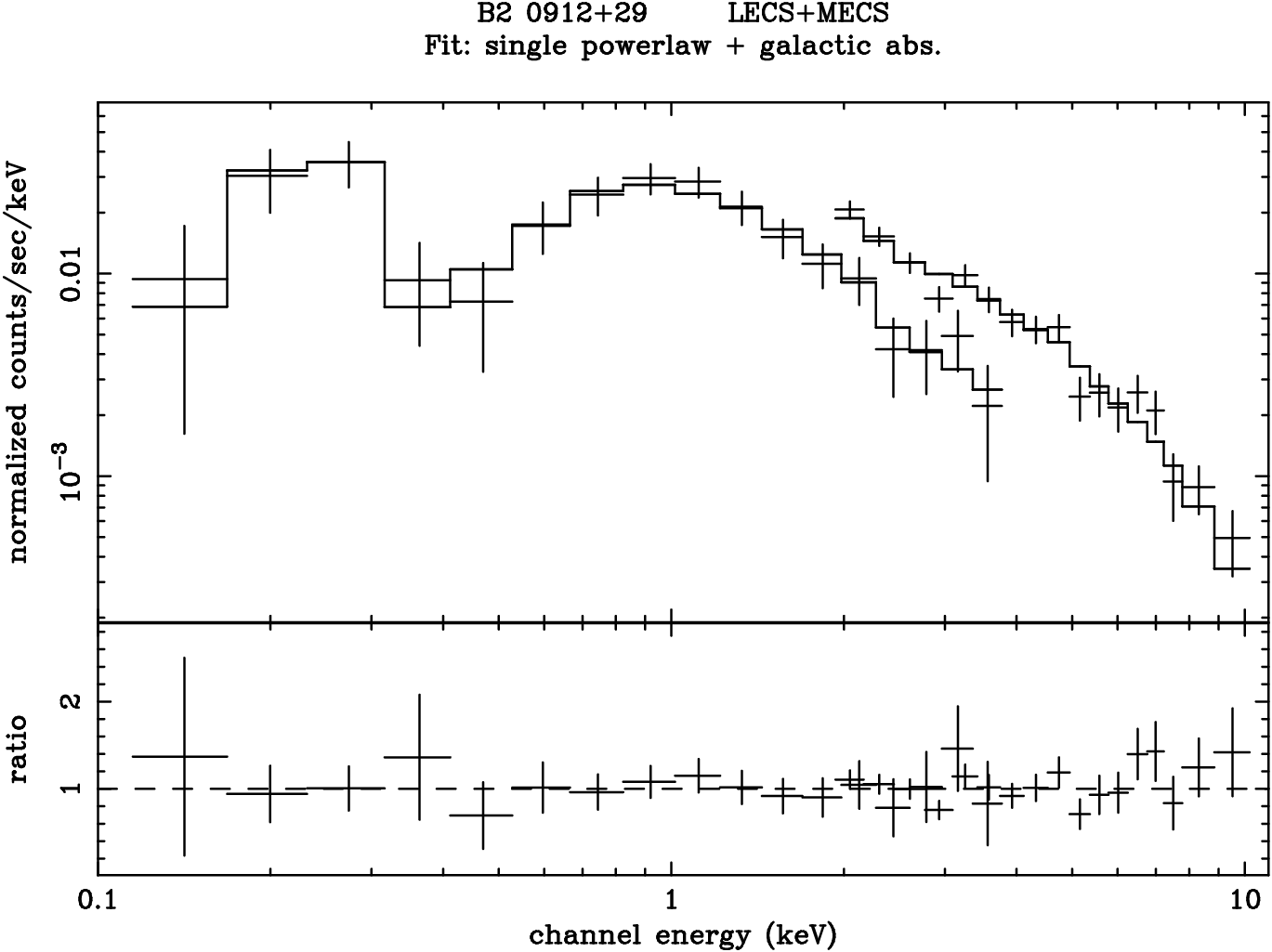,width=9cm,angle=-90}}
\end{figure}
\begin{figure}
\centerline{\psfig{figure=1144_gal_pow.ps,width=9cm,angle=-90}}
\medskip
\centerline{\psfig{figure=1519_gal_pow.ps,width=9cm,angle=-90}}
\medskip
\centerline{\psfig{figure=4c_5627_gal_pow.ps,width=9cm,angle=-90}}
\end{figure}
\begin{figure}
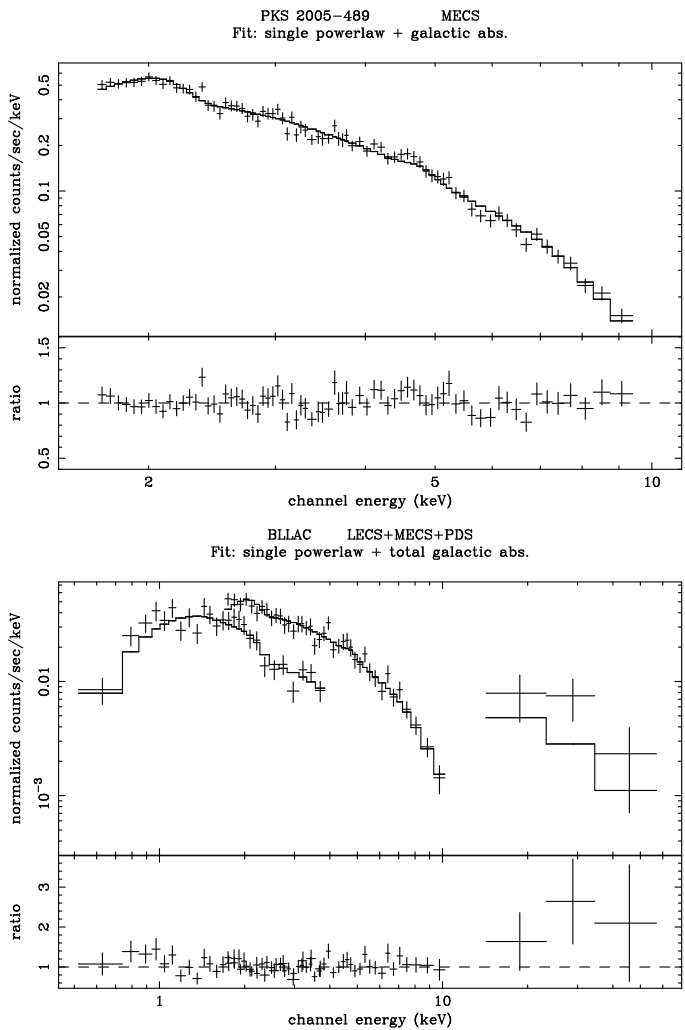

\centerline{\psfig{figure=2005_gal_pow.ps,width=9cm,angle=-90}}
\medskip
\centerline{\psfig{figure=bllac_lmp_gal_pow.ps,width=9cm,angle=-90}}
\caption{\sax data and fitted spectra for our sources, and ratio of data to
fit. Data are from the LECS and MECS instruments, apart from PKS 2005$-$489,
which only has MECS data, and BL Lac, which has also a PDS detection. The data
are fitted with a single power-law model with Galactic absorption.}
\end{figure}

\section{Spectral Fits}

The spectral analysis was performed with the XSPEC 10.0 package.
Using the program GRPPHA, the
spectra were rebinned with more than 20 counts in every new bin and 
using the rebinning files provided by the SDC. Various checks using different
rebinning strategies have shown that our results are independent of the
adopted rebinning within the uncertainties. The data were analyzed applying 
the Gehrels statistical weight (Gehrels 1986), in case the resulting 
net counts were below 20 (typically
12-15 in the low energy band of LECS). The LECS/MECS normalization
factor was left free to vary in the range 0.65--1.0, as suggested by SDC (see
Handbook for NFI spectral Analysis). The X-ray spectra of our 
sources are shown in Fig. 2. 

\subsection{Power-law Fits}

At first, we fitted the combined LECS and MECS data with a single power-law
model with Galactic and free absorption. The absorbing column was
parameterized in terms of N$_{\rm H}$, the HI column density, with heavier
elements fixed at solar abundances and cross sections taken from Morrison and
McCammon (1983). The Galactic value was derived from the {\tt nh} program at
HEASARC (based on Dickey \& Lockman 1990), when more accurate estimates were
not available (see Table 1). The Galactic \nh~value for BL Lac was fixed at
two values: that from Elvis, Lockman \& Wilkes (1989), based on dedicated 21
cm observations, and the sum of this value and that inferred from millimeter
observations, which include the contribution from molecular hydrogen (Lucas \&
Liszt 1993). The \nh~parameter was also set free to vary for all sources 
(apart from PKS 2005$-$489, which has no LECS data) to check for intrinsic  
absorption and/or indications of a ``soft-excess.''


Our results are presented in Table 3, which gives the name of the source in
column (1), \nh~in column (2), the energy index $\alpha_{\rm x}$ in column
(3), the 1 keV flux in $\mu$Jy in column (4), the unabsorbed $2 - 10$ keV and
$0.1 - 2.4$ keV fluxes in columns (5)-(6), the LECS/MECS normalization in
column (7), the reduced chi-squared and number of degrees of freedom,
$\chi^2_{\nu}/$(dof) in column (8), and the F--test probability that the
decrease in $\chi^2$ due to the addition of a new parameter (free \nh) is
significant in column (9). The errors quoted on the fit parameters are the
90\% uncertainties for one and two interesting parameters, for Galactic and
free \nh~respectively. The errors on the 1 keV flux reflect the statistical
errors only and not the model uncertainties. 
 
Two results are immediately apparent from Table 3. First, the fitted \nh~values
agree with the Galactic ones for most sources (within the rather large 
errors); this is confirmed by an
$F$-test which shows that the addition of \nh~as a free parameter does not
result in a significant improvement in the $\chi^2$ values (column 9 of Table
3), with the exception of BL Lac (F--test done for larger \nh~value). Second, 
the
fitted energy indices are relatively flat, \ax~$\la 1$ within the errors
for all but two sources. The
mean value is $\langle \alpha_{\rm x} \rangle = 0.98\pm0.10$ and the weighted
mean is $\langle \alpha_{\rm x} \rangle = 1.26\pm0.03$. This latter value is
clearly dominated by PKS 2005$-$489, an HBL with a very well determined
slope. Excluding this source and the other HBL, B2 0912+29, we derive a mean
value $\langle \alpha_{\rm x} \rangle = 0.87\pm0.09$ and a weighted mean
$\langle \alpha_{\rm x} \rangle = 0.83\pm0.09$.

Some of our sources appear to show a low-energy excess, as illustrated by the
fact that the best fit \nh~in Table 3 is below the Galactic value for four
sources, namely PKS 0048$-$097, PKS 1144$-$379, PKS 1519$-$273, and BL Lac. We
then fitted a broken power-law model to these data. Although this resulted in
a better fit, an F--test shows that the improvement is more suggestive than
significant, with probabilities ranging from $86-88\%$ for PKS 0048$-$097 and
PKS 1144$-$379, to $\sim 93\%$ for PKS 1519$-$273 and BL Lac. The best-fit
spectra, however, all point in the same direction: a flatter component
emerging at higher energies. In fact, they are all obviously concave with
quite a large spectral change, with $\langle \alpha_{\rm S} - \alpha_{\rm H}
\rangle =0.8\pm0.1$, and energy breaks around $E \sim 1 - 3$ keV. The hard
X-ray spectral index is $\langle \alpha_{\rm H} \rangle = 0.6\pm0.1$, while
$\langle \alpha_{\rm S} \rangle = 1.5\pm0.1$. Some evidence for concave
spectra comes also from the shape of the ratio of the data to the single
power-law fits, shown in Fig. 2. We note that by adding up the $\chi^2$ values
for the four sources, an F--test shows that the improvement in the fit
provided by a double power-law model {\it for the four sources together} is
significant at the $\sim 96\%$ level.

\begin{table*}
\begin{center}
{\bf Table 3. Single power-law fits, LECS + MECS}
\begin{tabular}{lllllllll}
\vspace*{-1mm}\\
\hline
\vspace*{-2mm} \\
Name  & N$_H$ & $\alpha_{x}$ & $F_{1keV}$ & $F_{[2-10]}$ & $F_{[0.1-2.4]}$ & 
Norm    & $\chi^2_{\nu}$/dof & F--test, notes \\
  & $10^{20}$ cm$^{-2}$ & & $\mu$Jy     & erg cm$^{-2}$ s$^{-1}$ & erg cm$^{-2}$ s$^{-1}$ & (Lecs/Mecs) & &  
fixed-free N$_{\rm H}$ 
\vspace*{1mm}\\
\hline
\vspace*{-3mm} \\
PKS 0048$-$097  & 3.85 fixed             & $0.9^{+0.4}_{-0.4}$ & $0.3^{+0.2}_{-0.1}$ & 1.45e-12 &
2.37e-12 & 0.82 & 0.58/12 &   \\
              & $2.0 (<18)$ & $0.9^{+0.5}_{-0.5}$ &   $0.3^{+0.2}_{-0.1}$ & 1.48e-12 &
2.11e-12 & 0.81 & 0.59/11 & 63\%
\vspace*{1mm}\\
\hline
\vspace*{-3mm} \\
OJ 287        & 2.75 fixed               & $0.6^{+0.2}_{-0.2}$ & $0.3^{+0.1}_{-0.1}$ & 2.46e-12 &
2.05e-12 & 0.66 & 1.00/9 &   \\
              & $21 (<163)$ & $0.7^{+0.6}_{-0.4}$ &  $0.4^{+0.4}_{-0.1}$ & 2.47e-12 &
2.65e-12 & 0.66 & 0.97/8  & 70\%
\vspace*{1mm}\\
\hline
\vspace*{-3mm} \\
B2 0912+29    & 2.11 fixed               & $1.3^{+0.1}_{-0.1}$ &$1.1^{+0.1}_{-0.1}$ & 2.86e-12 &
1.01e-11 & 0.74 & 0.76/38 &   \\
              & $2.8^{+1.6}_{-1.1}$ & $1.3^{+0.2}_{-0.2}$ &  $1.2^{+0.2}_{-0.2}$ & 2.80e-12 &
1.17e-11 & 0.73 & 0.74/37 & 86\%
\vspace*{1mm}\\
\hline
\vspace*{-3mm} \\
PKS 1144$-$379  & 7.64 fixed             & $0.6^{+0.3}_{-0.2}$ &$0.14^{+0.06}_{-0.04}$ & 1.00e-12 &
0.89e-12 & 0.91 & 1.03/17 &   \\
              & $0.0 (<21) $ & $0.6^{+0.3}_{-0.3}$ &
              $0.14^{+0.05}_{-0.03}$ & 0.99e-12 &
0.87e-12 & 0.86 & 0.98/16 & 82\%
\vspace*{1mm}\\
\hline
\vspace*{-3mm} \\
PKS 1519$-$273  & 8.66 fixed             & $1.1^{+0.4}_{-0.4}$ & $0.2^{+0.1}_{-0.1}$ & 0.62e-12 &
1.51e-12 & 0.77 & 0.67/8  &   \\
              & $5.2 (<44) $ & $1.0^{+0.6}_{-0.5}$ & $0.2^{+0.1}_{-0.1}$  & 0.63e-12 &
1.32e-12 & 0.76 & 0.74/7  & 36\%
\vspace*{1mm}\\
\hline
\vspace*{-3mm} \\
4C 56.27      & 4.16 fixed               & $1.1^{+0.3}_{-0.4}$ &$0.3^{+0.1}_{-0.1}$ & 0.98e-12 &
2.35e-12 & 0.71 & 0.49/7 &   \\
              & $5.7 (<75) $ & $1.1^{+0.7}_{-0.5}$ & $0.3^{+0.2}_{-0.1}$  & 0.98e-12 &
2.41e-12 & 0.72 & 0.57/6 & 10\%
\vspace*{1mm}\\
\hline
\vspace*{-2mm} \\
PKS 2005$-$489  & 5.08 fixed             & $1.33^{+0.04}_{-0.04}$ &  $25.4^{+1.1}_{-1.1}$  & 6.09e-11 &
2.60e-10 &  --- & 1.04/73 &  ---  \\
\vspace*{1mm}\\
\hline
\vspace*{-3mm} \\
BL Lac        & 20.15 fixed              & $0.83^{+0.08}_{-0.08}$ & $2.2^{+0.2}_{-0.2}$ & 1.12e-11 &
1.54e-11 & 0.72 & 0.92/63 & 21 cm maps \\
              & 36.0  fixed              & $0.96^{+0.08}_{-0.08}$ &   $2.7^{+0.3}_{-0.3}$ & 1.11e-11 &
2.03e-11 & 0.75 & 0.94/63 & H atom. + molec. \\
              & $27^{+10}_{-8}$   & $0.89^{+0.13}_{-0.13}$ &  $2.4^{+0.4}_{-0.3}$ & 1.12e-11 &
1.74e-11 & 0.73 & 0.89/62 & 96\% \\
              & 36.0 fixed        & $0.95^{+0.08}_{-0.08}$ &  $2.7^{+0.3}_{-0.3}$ & 1.12e-11 &
1.97e-11 & 0.75 & 0.96/66 & including PDS$^a$ \\
              & $26^{+10}_{-8}$   & $0.87^{+0.13}_{-0.13}$ & $2.4^{+0.3}_{-0.3} $ & 1.12e-11 &
1.67e-11 & 0.73 & 0.90/65 & including PDS$^a$ 
\vspace*{1mm}\\
\hline
\vspace*{-3mm} \\
\multicolumn{9}{l}{\footnotesize{$^a$ PDS maximum detection $3.8\sigma$ between 13 and 35 keV}}\\
\multicolumn{9}{l}{\footnotesize{Note: unless otherwise indicated, the errors 
are at $90\%$ confidence level for one (with fixed N$_{\rm H}$) and two 
parameters of interest.}} 
\end{tabular}
\end{center}
\end{table*}

\subsection{The PDS Detections}
Two sources were detected also by the PDS instrument: BL Lac and OJ 287.
Due to the low statistics, the spectra have been heavily rebinned, resulting
in three and one points for BL Lac and OJ 287 respectively, in the detection
range. The significance level of the detection, obtained grouping the
channels, is quite high for BL Lac ($3.8 \sigma$), while it is only marginal
for OJ 287 ($2.3 \sigma$). The relatively low flux for the latter source is at
the level expected from background fluctuations, and therefore we do not
regard this detection as real. 

Table 3 reports the results of a single power-law fits including the data from
all the three Narrow Field Instruments for BL Lac. The normalization factor
between PDS and MECS was fixed at 0.86, as derived from intercalibration tests
performed on known sources (see the SDC Handbook). The PDS points are
compatible with the best fit of the LECS and MECS data, even if slightly above
the model. 

Because of the absence of imaging capabilities, the PDS spectra can be 
contaminated by serendipitous sources in its field of view.
We therefore first checked the MECS image of BL Lac for the presence of other
sources, 
founding two in the field but with count rates a factor of 
10 and 40 fainter than the target.
However, given its wide field of view ($\sim1.4^{\circ}$ FWHM),
larger than the LECS and MECS ones ($\sim28^{\prime}$ for the MECS),
there is also the 
possibility of a contamination  by
hard serendipitous sources not
visible in the MECS images. We therefore checked WGACAT, the publicly
available database of ROSAT sources (White, Giommi, \& Angelini 1995), for
serendipitous X-ray sources within $\sim90^{\prime}$ from BL Lac. We found 
five,
but all of them were at least an order of magnitude fainter than our target,
and with relatively steep spectral indices (derived from the hardness
ratios). It then follows that it is very unlikely that any of these sources
can contribute significantly to the PDS flux of BL Lac, although the 
possibility remains that some hard X-ray sources might not have been detected 
by ROSAT. 

\subsection{Notes on individual sources}
{\bf PKS 1144--379.} LECS data below 1 keV present a clear trend, suggesting a steep
spectral index with stronger evidence than that provided by the F--test, given
the few points involved. MECS data present a feature around 4 keV but given the 
available S/N and resolution it is hard to assess the reality of this feature. 

{\bf OJ 287.} MECS data seem to show an emission feature around 2.3 keV but, again,
the low S/N does not allow reliable conclusions to be drawn.

The X-ray spectral index we obtain ($\alpha_{\rm x} = 0.6\pm0.2$)
agrees with the value derived by Kubo et al. (1998) of $0.62\pm0.01$,
based on ASCA observations made in November 1994. Our flux, however,
is $\sim 50\%$ smaller.

{\bf PKS 2005--489.} This source experienced a pronounced flare in November 1998,
about two years after our \sax~observations. Tagliaferri et al. (2001)
observed it with \sax~on November 1--2 and fitted a single power-law to the
data over the $0.1 - 200$ keV range with $\alpha_{\rm x} = 1.18\pm0.02$ and
free \nh. This is slightly flatter than our value of $1.33\pm0.04$, derived
between 1 and 10 keV (as we do not have LECS data) 
and assuming a Galactic \nh~close to their best-fit free
\nh~(see also Fig. 6). Their data actually require a broken power-law with a
break around 2 keV and a steepening $\Delta \alpha_{\rm x} \sim 0.2$ at higher
energies. Their X-ray flux was $\sim 3$ times higher than our value. Perlman
et al. (1999) followed the evolution of the flare between October 14 and
December 31 when the X-ray flux changed by a factor of four. The X-ray
spectral index in the $2-10$ keV band also varied between 1.3 and 1.8.

{\bf BL Lac.} Our results are consistent with those of Sambruna et al. (1999), based
on ASCA observations obtained in November 1995. Their single power-law fit,
assuming the same (fixed) \nh~value, has an energy index $\alpha_{\rm x} =
1.08\pm0.03$, to be compared with our value of $0.96\pm 0.08$. Their 1 keV
flux, derived for their fit with free \nh, and their broken power-law fit are
also consistent with ours. This shows that by the time of our
\sax~observations (Nov. 1997) the source had returned to its pre-flare
status. In July 1997, in fact, during its optical/X-ray/$\gamma$-ray flare,
the X-ray flux of BL Lac was $\sim 3$ times higher, with a much flatter X-ray
spectrum ($\alpha_{\rm x} \sim 0.4 - 0.7$; Tanihata et al. 2000; see also
Fig. 6). 

\section{ROSAT PSPC data}
In order to compare our results with previous (soft) X-ray observations and
especially to take advantage of the higher resolution and collecting area at
low energies, we used data from the ROSAT
Position Sensitive Proportional Counter (PSPC) public archive. 
The 1 Jy BL Lac ROSAT data had been originally published
by Urry et al. (1996), while those for B2 0912+29 were published by Lamer et
al.  (1996). In order to ensure a uniform procedure for the whole sample, we
have re-analyzed all ROSAT data, obtaining results consistent within the
errors with those already published.

The journal of the ROSAT observations is given in Table 4. The basic event
files from the archive have been corrected for gain variations on the detector
surface with the program PCSASSCORR in FTOOLS, when not already done by the
Standard Reduction process (SASS version 7\_8 and later, M. Corcoran, private
communication). Since all the sources were ROSAT targets a standard 
extraction radius of $3^{\prime}$ ($2.5^{\prime}$
when serendipitous sources were present in the field or when the source was
particularly weak) was used, to avoid the possible loss of soft photons due to
the ghost imaging effect. We have used the appropriate response matrices for
the different gain levels of the PSPC B detector before and after 14 Oct. 1991
(gain1 and gain 2, respectively). The background has been evaluated in two
circular regions (of radius $\sim 20-30$ pixels) away from the central region
and from other serendipitous sources, but inside the central rib ring of the
detector. The spectra have been rebinned (using GRPPHA) to have at least 20
net counts in every new bin, to justify the use of \chisq~statistics. Channel
1-11 and 212-256 have been excluded from the analysis, due to 
calibration uncertainties. 

As for the \sax~data, we fitted the ROSAT PSPC data with a single power-law
model with Galactic and free absorption. Our results are presented in Table 5,
which gives the name of the source in column (1), \nh~in column (2), the
energy index $\alpha_{\rm x}$ in column (3), the 1 keV flux in $\mu$Jy in
column (4), the unabsorbed $0.1 - 2.4$ keV flux in column (5), the reduced
chi-squared and number of degrees of freedom, $\chi^2_{\nu}/$(dof) in column
(6), and the observing date in column (7). The errors quoted on the fit
parameters are the 90\% uncertainties for one and two interesting parameters,
for Galactic and free \nh~respectively. The errors on the 1 keV flux reflect
the statistical errors only and not the model uncertainties. 

Table 5 shows that the fitted \nh~values are consistent with the Galactic ones
for most sources; this is confirmed by an $F$-test which shows that the
addition of \nh~as a free parameter does not result in a significant
improvement in the $\chi^2$ values. The fit for PKS 2005$-$489 is not
particularly good, especially for Galactic \nh. The fact that the 
free \nh~value is lower than the Galactic one suggests the presence of a 
``soft excess''. Indeed, Comastri et al. (1997) fitted a broken-power law
model to these data, with a steep soft index $\sim 4.7 - 4.8$. 
The fitted energy indices are
relatively steep (apart from 4C 56.27). The mean value is $\langle \alpha_{\rm
x} \rangle = 1.44\pm0.22$, the weighted mean is $\langle \alpha_{\rm x}
\rangle = 2.01\pm0.02$. This latter value is clearly dominated by PKS
2005$-$489, an HBL with a very well determined slope. Excluding this source
and 4C 56.27 we derive a mean value $\langle \alpha_{\rm x} \rangle =
1.51\pm0.14$, with a weighted mean $\langle \alpha_{\rm x} \rangle =
1.53\pm0.03$. 

\begin{table*}
{\bf Table 4. ROSAT journal of observations} \\
\begin{tabular}{lrcc}
\hline
Name & exposure  &  full band net count rate & Observing Date \\
     &   sec.    &    (cts/s) &    \\
\hline
PKS 0048$-$097& \dig8359 & $0.408\pm0.007$ & 1993 Jul 4-15 \\ 
OJ 287        & \dig3566 & $0.285\pm0.010$ & 1991 Apr 16 \\ 
              & \dig6702 & $0.602\pm0.010$ & 1991 Nov 10-11 \\ 
              & \dig3277 & $0.621\pm0.014$ & 1993 Oct 19 \\
B2 0912+29    & \dig2809 & $0.428\pm0.013$& 1991 Apr 24 - May 6 \\
PKS 1144$-$379& \dig7745 & $0.115\pm0.004 $ & 1993 Jul 7-8 \\
PKS 1519$-$273& \dig2548& $0.106\pm0.008 $ & 1993 Aug 17-18 \\ 
4C 56.27      & \dig5896& $0.138\pm0.005 $ & 1992 Jun 19-20 \\
PKS 2005$-$489& 11320 & $2.760\pm0.016 $ & 1992 Apr 27-29 \\
              & 11457 & $1.667\pm0.012$ & 1992 Oct 28 - Nov 1 \\
BL Lac        & \dig2167& $0.176\pm0.010 $   & 1992 Dec 22-23 \\
\hline 
\end{tabular}
\end{table*}

\begin{table*}
{\bf Table 5. ROSAT PSPC, single power-law fits}
\begin{tabular}{lllllll}
\vspace*{-1mm}\\
\hline
\vspace*{-2mm} \\
Name  & N$_{\rm H}$ & $\alpha_{\rm x}$ & $F_{1keV}$ & $F_{[0.1-2.4]}$ & $\chi^2_{\nu}$/dof&  Observing Date \\
      & $10^{20}$ cm$^{-2}$ & & $\mu$Jy     & erg cm$^{-2}$ s$^{-1}$ &  &   
\vspace*{1mm}\\
\hline
\vspace*{-3mm} \\
PKS 0048$-$097 & 3.85 fixed             & $1.63\pm0.04$ & $0.86\pm0.03$ & 1.22e-11 & 0.76/55 & 1993 Jul 4-15 \\
               & $4.39^{+0.60}_{-0.58}$ & $1.79\pm0.19$ & $0.87\pm0.04$ & 1.52e-11 & 0.70/54 & 
\vspace*{1mm}\\
\hline
\vspace*{-3mm}\\
OJ 287        & 2.75 fixed             & $1.17^{+0.08}_{-0.08}$ & $0.61^{+0.04}_{-0.04}$ & 5.34e-12 & 0.79/31 & 1991 Apr 16 \\
              & $2.71^{+1.06}_{-0.92}$ & $1.15^{+0.36}_{-0.33}$ & $0.61^{+0.05}_{-0.05}$ & 5.27e-12 & 0.82/30 &         
\vspace*{3mm}\\
              & 2.75 fixed             & $1.62^{+0.04}_{-0.04}$ & $0.97^{+0.04}_{-0.04}$ & 1.34e-11 & 0.89/80 & 1991 Nov 10-11 \\
              & $2.37^{+0.44}_{-0.41}$ & $1.47^{+0.17}_{-0.17}$ & $0.96^{+0.04}_{-0.04}$ & 1.14e-11 & 0.86/79 &                       
\vspace*{3mm}\\
              & 2.75 fixed             & $1.29^{+0.07}_{-0.07}$ & $1.24^{+0.08}_{-0.08}$ & 1.22e-11 & 0.99/54 & 1993 Oct 19 \\ 
              & $2.72^{+0.67}_{-0.62}$ & $1.28^{+0.23}_{-0.23}$ & $1.24^{+0.08}_{-0.08}$ & 1.21e-11 & 1.01/53 &         
\vspace*{1mm}\\
\hline
\vspace*{-3mm} \\
B2 0912+29    & 2.11  fixed            & $1.53^{+0.08}_{-0.08}$ & $0.60^{+0.05}_{-0.05}$ & 7.46e-12 & 0.75/31 & 1991 Apr 24 - May 6 \\
              & $1.64^{+0.78}_{-0.64}$ & $1.34^{+0.33}_{-0.29}$ & $0.59^{+0.05}_{-0.05}$ & 6.05e-12 & 0.72/30 & 
\vspace*{1mm}\\
\hline
\vspace*{-3mm} \\
PKS 1144$-$379  & 7.64 fixed           & $1.37^{+0.14}_{-0.14}$ & $0.41^{+0.03}_{-0.03}$ & 4.41e-12 & 1.04/16 & 1993 Jul 7-8 \\
              &  $10.0^{+3.9}_{-2.3}$  & $1.71^{+0.41}_{-0.23}$ & $0.45^{+0.05}_{-0.04}$ & 7.08e-12 & 0.78/15 & 
\vspace*{1mm}\\
\hline
\vspace*{-3mm} \\
PKS 1519$-$273  & 8.66 fixed           & $1.12^{+0.26}_{-0.29}$ & $0.40^{+0.05}_{-0.05}$ & 3.35e-12 & 1.56/10 & 1993 Aug 17-18 \\
              & $32.7^{+55.5}_{-22.3}$ & $2.86^{+3.37}_{-1.63}$ & $0.84^{+1.71}_{-0.80}$ & 7.91e-11 & 0.95/9  & 
\vspace*{1mm}\\
\hline
\vspace*{-3mm} \\
4C 56.27      & 4.16   fixed         & $0.23^{+0.15}_{-0.15}$ & $0.45^{+0.03}_{-0.03}$ & 2.51e-12 & 0.79/23 & 1992 Jun 19-20 \\
              & $5.63^{+3.41}_{-1.99}$ & $0.46^{+0.40}_{-0.40}$ & $0.47^{+0.05}_{-0.05}$ & 2.79e-12 & 0.52/22 & 
\vspace*{1mm}\\
\hline
\vspace*{-2mm} \\
PKS 2005$-$489  & 5.08 fixed          & $2.25^{+0.02}_{-0.02}$ & $5.15^{+0.07}_{-0.07}$ & 1.74e-10 & 2.15/89 & 1992 Apr 27-29 \\
               & $4.28^{+0.19}_{-0.19}$& $1.99^{+0.06}_{-0.06}$ & $5.12^{+0.07}_{-0.07}$ & 1.18e-10 & 1.25/88 &          
\vspace*{3mm}\\
               &  5.08 fixed           & $2.43^{+0.02}_{-0.02}$ & $2.70^{+0.05}_{-0.05}$ & 1.22e-10 & 3.48/65 & 1992 Oct 28 - Nov 1\\
               & $3.54^{+0.23}_{-0.23}$& $1.91^{+0.08}_{-0.08}$ & $2.71^{+0.05}_{-0.05}$ & 5.54e-11 & 0.79/64 &        
\vspace*{1mm}\\
\hline
\vspace*{-3mm} \\
BL Lac        & 36.0   fixed           & $2.13^{+0.36}_{-0.36}$ & $1.55^{+0.14}_{-0.14}$ & 4.38e-11 & 0.56/11 & 1992 Dec 22-23 \\
              & $33.7^{+4.58}_{-2.44}$ & $2.01^{+2.35}_{-1.49}$ & $1.45^{+1.50}_{-0.62}$ & 3.43e-11 & 0.62/10 &            
\vspace*{1mm}\\
\hline
\vspace*{-3mm} \\
\multicolumn{7}{l}{\footnotesize{Note: the errors are at $90\%$ confidence 
level for one 
(with fixed N$_{\rm H}$) and two parameters of interest.}} \\
\end{tabular}
\end{table*}

\subsection{Comparison between \sax and ROSAT results}

Figure 3 shows the \sax 1 keV flux versus the corresponding ROSAT flux. The
errors reflect statistical uncertainties only and do not include model
uncertainties. Our sources display mild X-ray variability: the median value of
$f_{\rm \sax}/f_{\rm ROSAT}$ is 0.6 (0.5 excluding PKS 2005$-$489 which has
the largest value of this ratio, $\sim 5$). Fig. 3 should be compared with
Fig. 2 in Wolter et al. (1998), which shows the same plot for a sample of
eight HBL. There the two fluxes are within 30\% for most sources and the
points follow more closely the line of equal fluxes. We note that the 1 keV
flux has a very strong model dependence. We therefore evaluated the X-ray flux
ratio also in the $0.1-2.4$ keV range, a band common to both instruments and
less model dependent. The median value in this case is not very different,
$f_{\rm \sax}/f_{\rm ROSAT} = 0.4$. 

\begin{figure}
\centerline{\psfig{figure=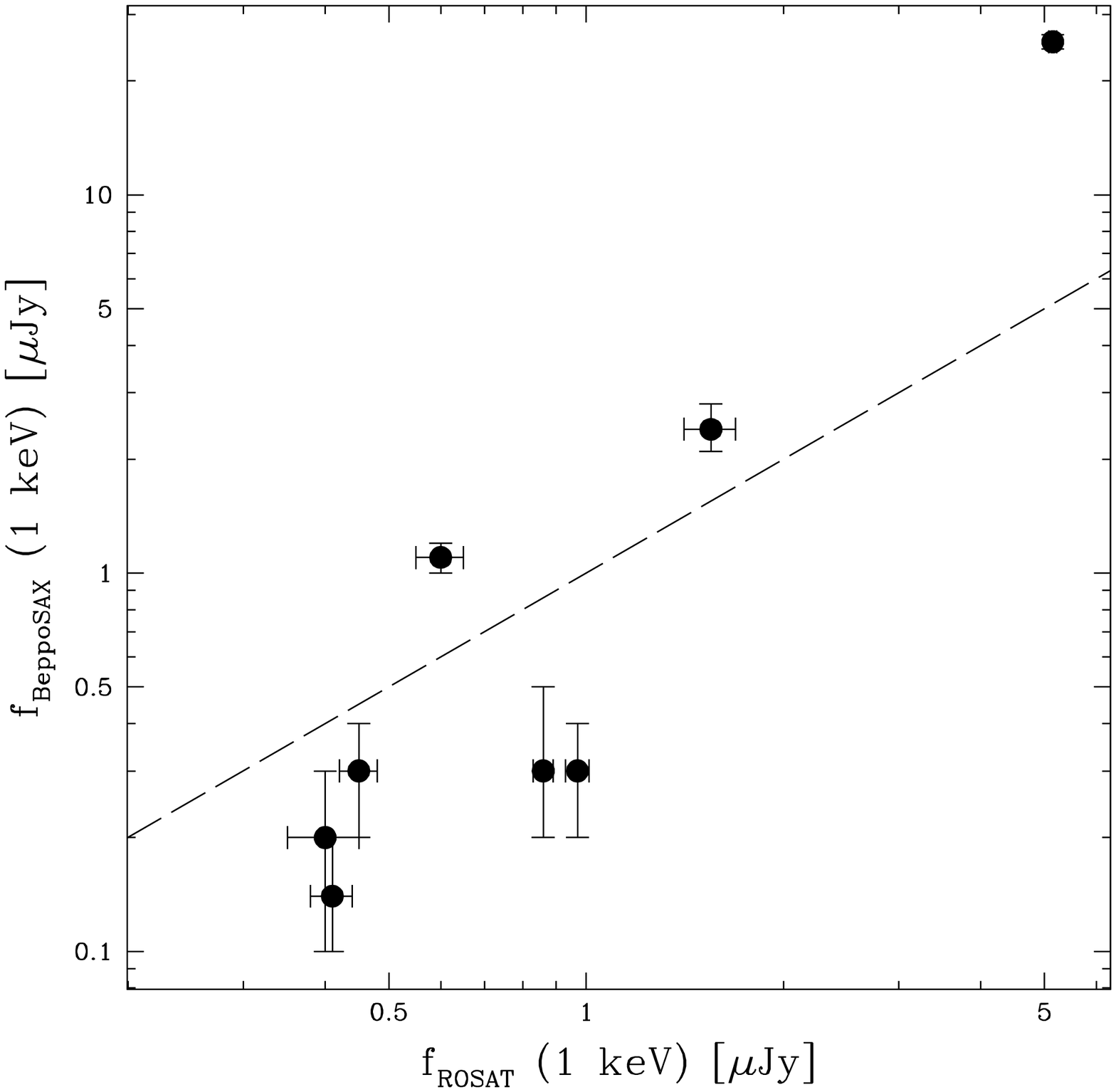,width=9cm}}
\caption{The 1 keV X-ray flux from our \sax data vs. the 1 keV X-ray
flux from ROSAT data. The dashed line represents the locus of $f_{\sax}=
f_{ROSAT}$. The errors reflect the statistical errors only and not the model
uncertainties. For OJ 287 and PKS 2005$-$489, which have multiple ROSAT 
observations, we took the observation with the largest X-ray flux 
(Oct. 1993 and Apr. 1992 respectively).}
\end{figure}

Figure 4 shows the \sax spectral index ($0.1-10$ keV) vs. the ROSAT spectral
index ($0.1-2.4$ keV). The larger \sax error bars for most of our sources, as
compared to ROSAT, are due to the worse photon statistics. (The PSPC count
rates, in fact, are typically a factor of 10 larger than the LECS ones.) All
but one source occupy the region of the plot where $\alpha_{\rm x}(\sax) \le
\alpha_{\rm x}(ROSAT)$. The interpretation of this plot is complicated by
variability effects, which affect the shape of the X-ray spectrum, and 
possibly by ROSAT miscalibrations (e.g., Iwasawa,
Fabian \& Nandra 1999; Mineo et al. 2000). However, a few points can be 
made. Again, as before,
the figure suggests a concave overall X-ray spectrum for our sources, with a
flatter component emerging at higher energies. We find $\alpha_{\rm x}(ROSAT)
- \alpha_{\rm x}(\sax) = 0.47\pm0.23$ (excluding the two HBL 
this becomes $\alpha_{\rm x}(ROSAT) - \alpha_{\rm x}(\sax) =
0.43\pm0.30$). This difference cannot be attributed to 
miscalibration effects alone which, if present, should steepen the ROSAT
slopes by $\sim 0.2 - 0.3$ (Mineo et al. 2000).

Again, this figure should be compared with Fig. 3 of Wolter et
al. (1998), which shows the same plot for a sample of 8 HBL. In that
case the \sax and ROSAT spectral indices agree within the errors for
all but one source.
 
\begin{figure}
\centerline{\psfig{figure=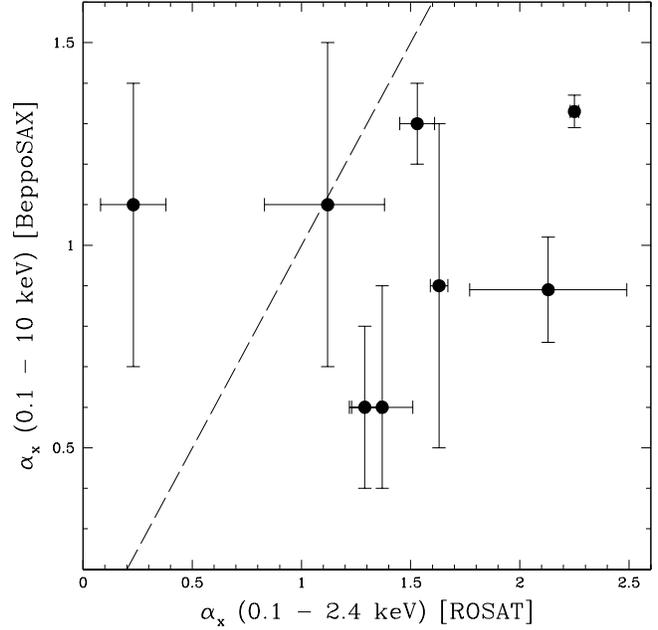,width=9cm}}
\caption{The \sax spectral index ($0.1-10$ keV) vs. the ROSAT spectral index
($0.1-2.4$ keV). The dashed line represents the locus of $\alpha_{\rm
x}(\sax)= \alpha_{\rm x}(ROSAT)$. For OJ 287 and PKS 2005$-$489, which have
multiple ROSAT observations, we took the observation with the largest X-ray 
flux (Oct. 1993 and Apr. 1992 respectively).}
\end{figure}

\section{Spectral Energy Distributions} 
To address the relevance of our \sax data in terms of emission processes in BL
Lacs, we have assembled multifrequency data for all our sources. The main
source of information was NED, and so most data are not simultaneous with our
\sax observations. For five of our sources, however, we were able to find
nearly-simultaneous (within a month) radio observations in the University of
Michigan Radio Astronomy Observatory (UMRAO) database. These are reported in
Table 6, which also gives the nearly-simultaneous radio-X-ray spectral index,
$\alpha_{\rm rx}$ with its error. This is defined between the rest-frame
frequencies of 4.8 GHz and 1 keV, and has been K-corrected using the X-ray
spectral indices given in Table 3 and radio spectral indices between 4.8 and
8.0 GHz derived from the UMRAO data ($z=0.3$ was assumed for the two sources
in the table without redshift information). Three of our sources were also
detected by EGRET so their energy distributions reach $\sim 5 \times 10^{24}$
Hz. The EGRET data come from the compilation of Lin et al. (1999), which
include the first entries in the Third EGRET Catalog.

The spectral energy distributions (SEDs) for our sources are shown in Figures
5 and 6, where filled circles indicate \sax data and the nearly-simultaneous
radio data and open symbols typically represent non-simultaneous literature
(NED) data. The \sax data have been converted to $\nu f_{\rm \nu}$ units using
the XSPEC unfolded spectra after correcting for absorption. ROSAT (from Tab.
5) and EGRET data are shown by a bow-tie that represent the spectral index
range.

We have fitted a homogeneous, one--zone synchrotron inverse Compton model to
the SED of our sources. The model is very similar to the one described in
detail in Spada et al. (2001; it is the ``one--zone'' version of it). It
assumes that the source is cylindrical, of size $R$ and width $\Delta
R^\prime=R/\Gamma$ (in the comoving frame, where $\Gamma$ is the bulk Lorentz
factor). The particle distribution $N(\gamma)$ is assumed to have the slope
$n$ [$N(\gamma)\propto \gamma^{-n}$] above the random Lorentz factor
$\gamma_{\rm c}$, where radiative losses dominate over adiabatic losses.  The
electron distribution is assumed to cut-off abruptly at $\gamma_{\rm max} >
\gamma_{\rm c}$. Below $\gamma_{\rm c}$ there can be two cases, depending on
the values of $\gamma_c$ and $\gamma_{\rm min}$. If $\gamma_{\rm
c}>\gamma_{\rm min}$, we have $N(\gamma) \propto \gamma^{-n+1}$ between
$\gamma_{\rm min}$ and $\gamma_{\rm c}$ and $N(\gamma)\propto \gamma^{-1}$
below $\gamma_{\rm min}$. Alternatively, if $\gamma_{\rm c} < \gamma_{\rm
min}$, then $N(\gamma) \propto \gamma^{-2}$ between $\gamma_{\rm c}$ and
$\gamma_{\rm min}$ and $\gamma^{-1}$ below $\gamma_{\rm c}$. According to
these assumptions, the random Lorentz factor of the electrons emitting most of
the radiation (i.e., emitting at the peaks of the SEDs), $\gamma_{\rm peak}$,
is determined by the relative importance of the adiabatic versus radiative
losses and can assume values in the range $\gamma_{\rm min}$--$\gamma_{\rm
max}$.

Photons produced externally to the jet (e.g., by the broad line region; BLR)
are considered only if they improve the fit. We account for them assuming that
a fraction 0.1 of the disk luminosity $L_{\rm disk}$ is reprocessed into line
emission by the BLR assumed to be located at $R_{\rm BLR}$. The source is
assumed to emit an intrinsic luminosity $L^\prime$ and to be observed with the
viewing angle $\theta$. The input parameters are listed in Table 7, which
gives the name of the source in column (1), $L^\prime$ in column (2), $L_{\rm
disk}$ in column (3), $R_{\rm BLR}$ in column (4), the magnetic field $B$ in
column (5), the size of the region $R$ in column (6), the Lorentz factor
$\Gamma$ in column (7), the angle $\theta$ in column (8), the slope of the
particle distribution $n$ in column (9), and finally $\gamma_{\rm min}$ and
$\gamma_{\rm peak}$ in columns (10) and (11) respectively. Note that $\gamma_{\rm peak}$ 
is a derived quantity and not an input parameter.

In the case of a pure synchrotron self--Compton model, all the above
parameters are constrained in sources for which: 1) we have an estimate of the
minimum timescale of variability; 2) both the synchrotron and the self--Compton
peak are well defined; 3) the spectral slopes before and above the peaks are
known; 4) the redshift is known. As discussed in Tavecchio, Maraschi \&
Ghisellini (1998), this suffices to fix the values of the magnetic field, the
intrinsic power of the source, the slopes of the emitting electron
distribution, the relativistic Doppler factor, and the dimension of the
source. When the radiation produced externally to the jet is important there
is one unconstrained unknown, but the superluminal motion of the radio knots
observed for many of these sources indicates values of the bulk Lorentz factor
in the range 10--15 on average, and we therefore use these values for our fits
(see, e.g., Ghisellini et al. 1998).

For the sources in our sample, we rarely have complete information
about the high-energy peak (we often have only an upper limit),
so we lack the direct determination of $\gamma_{\rm peak}$
and have only a limit on the determination of the magnetic field.

But in the model we use here the adiabatic losses play a crucial role, and
(within this model) we have an additional constrain with respect to the
simplest synchrotron inverse Compton model. Namely, the peak of the
synchrotron emission is either due to the electrons injected with $\gamma_{\rm
min}$, or it is due to the electrons for which adiabatic and radiative losses
balance.  This second condition (coupled with the value of the synchrotron
peak frequency) allows to estimate the value of the magnetic field, since for
our sources the synchrotron energy losses are important (as shown by the upper
limits in the EGRET energy range indicating an inverse Compton emission not
widely dominant).

A further constraint applies to sources in which external radiation could be
important.  In our model these external seed photons are thought to be
produced by the broad line region, reprocessing a fixed amount of the ionizing
flux produced by the accretion disk.  Therefore the accretion disk luminosity
has not to exceed the observed optical--UV continuum (for our sources we do
not have any evidence for the presence of a blue bump), nor the emission line
luminosities have to exceed the observed values.

The model fits are shown in Fig. 5 and 6 as solid (and dashed) lines. The
applied model is aimed at reproducing the spectrum originating in a limited
part of the jet, thought to be responsible for most of the emission. This
region is necessarily compact, since it must account for the fast variability
shown by all blazars, especially at high frequencies. The radio emission from
this compact regions is strongly self--absorbed, and the model cannot account
for the observed radio flux. This explains why the radio data are 
systematically above the model fits in the figures. 

For some sources the model fits present a complex behavior at $\gamma$--ray
energies (i.e., two peaks at high energies besides the synchrotron peak at
lower frequencies). In all such cases the first high energy peak is due to
synchrotron self--Compton emission, while the peak at the highest energies is
due to inverse Compton off external photons.

As shown in Table 7, the intrinsic luminosities, the source dimensions, the
bulk Lorentz factors and the viewing angles are quite similar for all sources,
while the magnetic field varies from 0.8 to 6 Gauss (with the smallest 
values corresponding to the least powerful sources). The need for
external seed photons for some sources, while indicative of a broad line
region, is not extremely compelling, since the required disk luminosities are
much smaller than those required in radio--loud quasars (see, e.g., Ghisellini
et al. 1998). The main difference between sources are in the derived value of
$\gamma_{\rm peak}$ with HBL having, not surprisingly, the larger values.
Moreover, $\gamma_{\rm peak}$ strongly (anti--)correlates with $L^\prime$,
that is powerful sources have the smallest values of $\gamma_{\rm peak}$, in
agreement with what previously found by Ghisellini et al. (1998). 

\begin{figure*}
\centerline{\psfig{figure=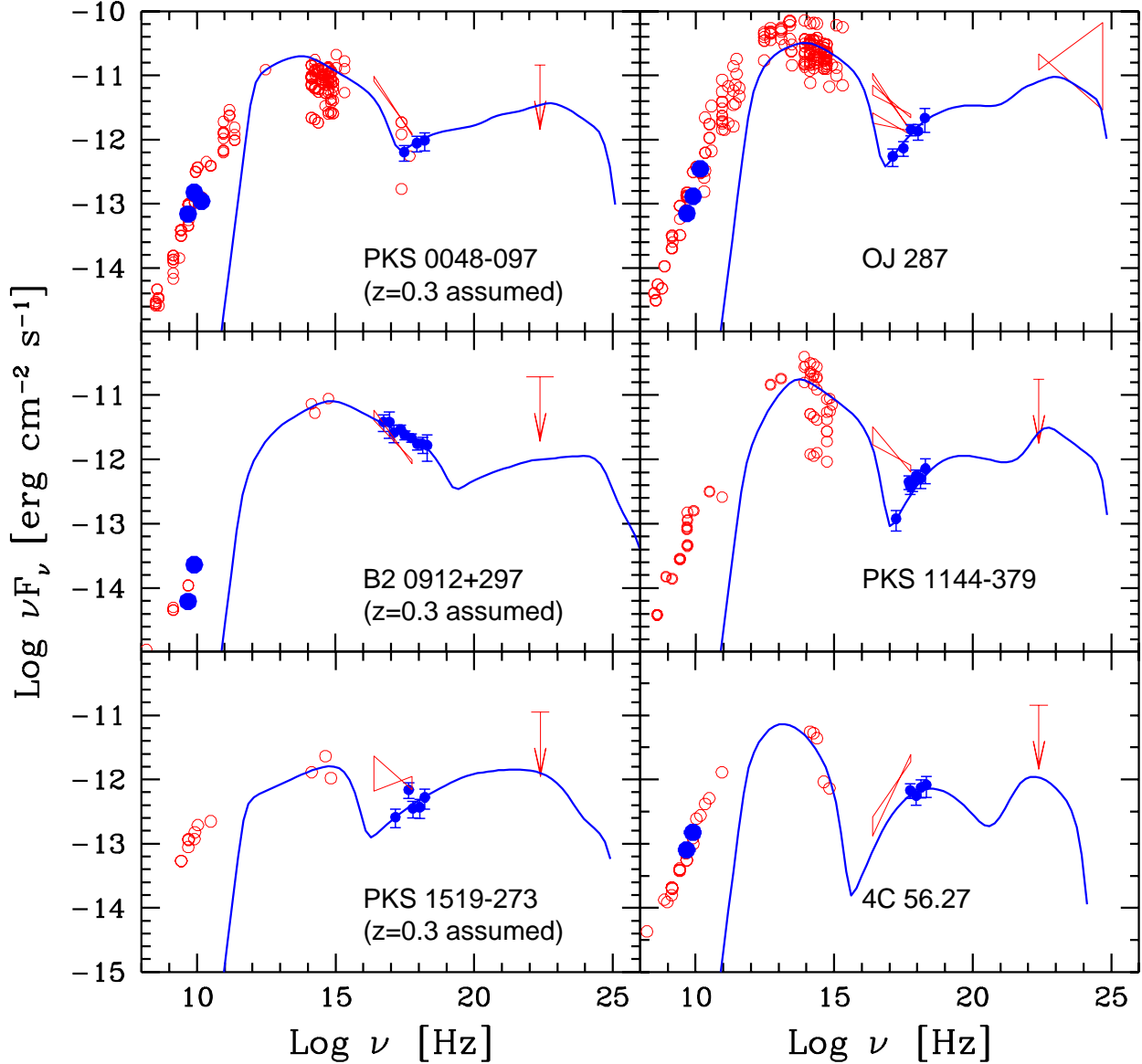,width=17cm}}
\vskip -1. true cm
\caption{Spectral energy distributions of six of our sources. Filled symbols
indicate \sax~data and nearly-simultaneous radio data from UMRAO, while open
symbols indicate data from the literature (NED). The solid lines correspond to
the one--zone homogeneous synchrotron and inverse Compton model calculated as
explained in the text, with the parameters listed in Table 7.}
\end{figure*} 

\begin{figure*}
\centerline{\psfig{figure=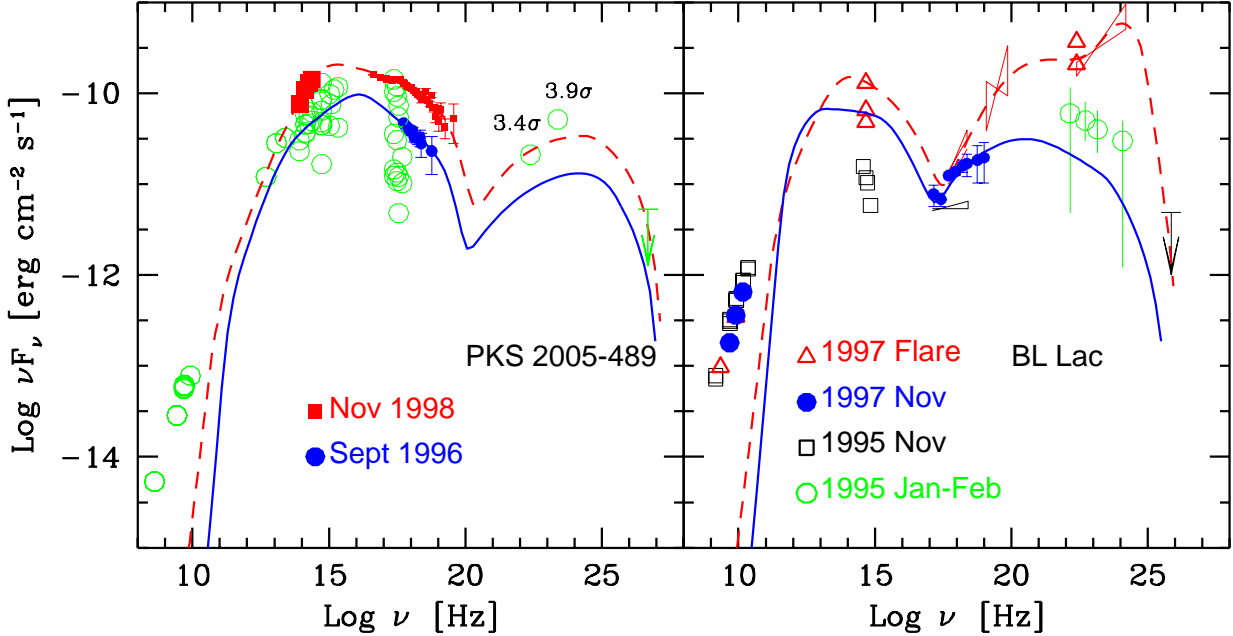,height=17cm}}
\vskip -8 true cm
\caption{Spectral energy distributions of PKS 2005$-$489 and BL Lac. PKS
2005$-$489: filled circles indicate our \sax~data (Sept. 1996), while filled
squares represent the \sax~observations of November 1998, with simultaneous
infrared data (Tagliaferri et al. 2001); open symbols indicate data from the
literature (NED). BL Lac: filled circles indicate our \sax~data (Nov. 1997)
and nearly-simultaneous radio data from UMRAO, other points correspond to
different observing dates, with triangles indicating the summer 1997 flare
(see Sambruna et al. 1999 for details on the various data). The lines (solid:
our data; dashed: other campaigns) correspond to the one--zone homogeneous
synchrotron and inverse Compton model calculated as explained in the text,
with the parameters listed in Table 7.}
\end{figure*} 

\begin{table*}
{\bf Table 6. Nearly-simultaneous Radio Observations}
\begin{tabular}{lrlrlrlc}
\hline
Name &F$_{\rm 4.8GHz}$&Observing date&F$_{\rm 8.0GHz}$&Observing date&F$_{\rm 14.5GHz}$&Observing date&$\alpha_{\rm rx}$ \\
     &              Jy&              &              Jy&              &           Jy& & \\
\hline
PKS 0048$-$097&$1.43\pm0.04$&1997 Dec 3 &$1.83\pm0.19$&1997 Dec 6 &$0.79\pm0.06$&1997 Dec 16&$0.85\pm0.03$\\
OJ 287        &$1.47\pm0.05$&1997 Nov 15&$1.63\pm0.08$&1997 Nov 25&$2.44\pm0.05$&1997 Dec 17&$0.86\pm0.02$\\
B2 0912+29    &$0.13\pm0.03$&1997 Dec 2 &$0.29\pm0.04$&1997 Dec 8 & ... & ...&$0.62\pm0.01$\\
4C 56.27      &$1.66\pm0.07$&1997 Oct 18&$1.87\pm0.09$&1997 Oct 7 & ... & ...&$0.84\pm0.02$ \\
BL Lac        &$3.73\pm0.13$&1997 Nov 9 &$4.53\pm0.12$&1997 Nov 21&$4.47\pm0.12$&1997 Nov 13&$0.80\pm0.01$\\
\hline
\end{tabular}
\end{table*}
\begin{table*}
{\bf Table 7. Model Input Parameters }
\begin{center}
\begin{tabular}{lllllllllll}
\hline
Name  &$L^\prime$       &$L_{\rm disk}$   &$R_{\rm BLR}$ &$B$ &$R$    &$\Gamma$ &$\theta$ &$n$  &$\gamma_{\rm min}$ &$\gamma_{\rm peak}$ \\
      & erg s$^{-1}$    &erg s$^{-1}$ & cm       &G   & cm    &         &         &     &     &        \\
\hline
PKS 0048$-$097 &3.0e42  &1e45         &6e17      &5   &1.6e16 &12       &4.5      &3.5  &50   &560     \\
OJ 287         &3.0e42  &1e45         &5e17      &5   &1.3e16 &13       &4        &3.5  &150  &660     \\
B2 0912+29     &8.0e41  &1e44         &5e17      &3   &1.2e16 &13       &4        &3.45 &200  &2300     \\   
PKS 1144$-$379 &6.0e42  &6e44         &5e17      &6   &2.0e16 &15       &3        &3.6  &400  &400      \\
PKS 1519$-$273 &5.0e41  &...          &...       &0.8 &1.0e16 &13       &4.7      &3.6  &40   &10,000   \\ 
4C 56.27       &6.0e42  &5e44         &5e17      &5   &2.0e16 &13       &5        &3.95 &200  &500       \\
PKS 2005$-$489 &2.0e41  &...          &...       &1.5 &1.0e16 &13       &3        &3.5  &600  &12,000    \\
BL Lac         &1.1e42  &...          &...       &1.2 &1.5e16 &10       &5.       &4.05  &300   &300     \\    
\hline 
\end{tabular}
\end{center}
\end{table*}

\section{X-ray spectral index and the synchrotron peak frequency}

One of the aims of this project was to study the dependence of the X-ray
spectral index on the synchrotron peak frequency $\nu_{\rm peak}$ found by
Padovani \& Giommi (1996) and Lamer et al. (1996) from ROSAT data by using the
broader \sax energy band. Padovani \& Giommi (1996) found a strong
anti-correlation between $\alpha_{\rm x}$ and $\nu_{\rm peak}$ for HBL (i.e.,
the higher the peak frequency, the flatter the spectrum), while basically no
correlation was found for LBL. This was interpreted as due to the tail of the
synchrotron component becoming increasingly dominant in the ROSAT band as
$\nu_{\rm peak}$ moves closer to the X-ray band (see Fig. 7 of Padovani \&
Giommi 1996).

The \sax version of this dependence is shown in Figure 7, which plots the
X-ray spectral index ($0.1-10$ keV) vs. the logarithm of the peak frequency 
for
our sources (filled circles), the HBL studied by Wolter et al. (1998; open
squares), and other BL Lacs studied by {\it BeppoSAX}. These include, in order
of increasing peak frequency: ON 231 (star; $\alpha_{\rm x}$ in the $0.1-3.8$
keV range; Tagliaferri et al. 2000), S5 0716+714 (open circle; Giommi et
al. 1999), PKS 2155$-$304 (filled triangle; Giommi et al. 1998), MKN 421
(cross; $\alpha_{\rm x}$ in the $0.1-1.6$ keV range; Fossati et al. 2000), 1ES
2344+514 (open triangle; Giommi, Padovani \& Perlman 2000), and MKN 501
(filled square; Pian et al. 1998). When more than a value of $\alpha_{\rm x}$
was available for these BL Lacs we picked the one corresponding to the largest
$\nu_{\rm peak}$. The $\nu_{\rm peak}$ values for the sources studied in this
paper have been taken from Sambruna, Maraschi \& Urry (1996), who fitted a
parabola to the $\nu f_{\rm \nu}$ broad-band spectra, except for B2 0912+29
and PKS 2005$-$489. For these two sources we derived $\nu_{\rm peak}$ as
described in Padovani \& Giommi (1996). The former source, in fact, was not
included in the sample studied by Sambruna et al. (1996), while for the latter
the derived value was clearly too high (see, e.g., Comastri, Molendi \&
Ghisellini 1995). The $\nu_{\rm peak}$ values for the HBL studied by Wolter et
al. (1998) are taken from that paper and similarly the values for the
additional sources are normally taken from the referenced papers.

Fig. 7, although with less statistics, basically confirms the ROSAT findings,
namely a strong anti-correlation between $\alpha_{\rm x}$ and $\nu_{\rm peak}$
for HBL and no correlation for LBL, with an initial increase in 
$\alpha_{\rm x}$ going from LBL to HBL. A few differences, however, are worth
mentioning. First, the range in $\alpha_{\rm x}$ is somewhat smaller ($\sim 1$
vs. $\sim 3$). This is likely due to the larger energy range over which
$\alpha_{\rm x}$ is measured ($0.1-10$ keV for \sax vs. $0.1-2.4$ keV for
ROSAT). Objects with very steep ROSAT $\alpha_{\rm x}$, in fact, are those in
which synchrotron emission is nearing the exponential cut-off; by having a
larger band \sax includes flatter, higher energy emission due to inverse
Compton. Second, the wide $0.1-100$ keV coverage of \sax has allowed the
detection of spectacular spectral variability with $\nu_{\rm peak}$ reaching
$\ga 10$ keV. As predicted by Padovani \& Giommi (1996), these sources display
very flat $\alpha_{\rm x}$ ($\sim 0.5 - 0.8$), since \sax is sampling the 
peak 
of the synchrotron emission. Note that in this case the flat X-ray spectrum
is {\it not} associated with inverse Compton emission, although extreme
HBL (objects to the far right in Fig. 7) have synchrotron X-ray spectra as 
flat as extreme LBL (objects to the far left of the figure). 
In other words, there are two very different mechanisms which can produce a
flat X-ray spectrum in BL Lacs: inverse Compton emission or synchrotron
emission with peak frequency in the hard X-ray band.

\begin{figure}
\centerline{\psfig{figure=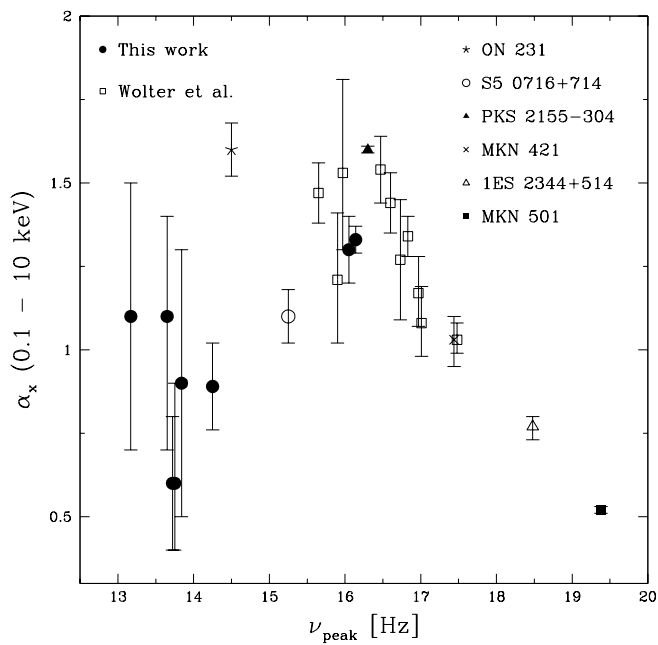,width=9cm}}
\caption{The \sax~spectral index ($0.1-10$ keV) vs. the logarithm of the peak
frequency for our sources (filled circles), the HBL studied by Wolter et
al. (1998; open squares), ON 231 (star), S5 0716+714 (open circle), PKS
2155$-$304 (filled triangle), MKN 421 (cross), 1ES 2344+514 (open triangle),
and MKN 501 (filled square). See text for details and references.}
\end{figure}

\section{Results and conclusions}

We have presented new \sax observations of eight BL Lacertae objects, all but
one selected from the 1 Jy sample. Six of our sources are LBL, i.e., they are
characterized by a peak in their multifrequency spectra at infrared/optical
energies. A relatively simple picture comes out from this paper: a dominance
of inverse Compton emission in the X-ray band of LBL, with $\sim 2/3$ of the
sources showing also a likely synchrotron component. This result rests on
various pieces of evidence:

\begin{itemize}

\item the relatively flat ($\alpha_{\rm x} \sim 0.9$) \sax spectra of our LBL
sources (Tab. 3). Moreover, single power-law fits to the \sax data
show a best fit $N_{\rm H}$ below the Galactic value for four out of six of 
our LBL, while
residuals to the single power-law fits with Galactic $N_{\rm H}$ also show
evidence for concave spectra (Fig. 2). Broken power-law models improve the
fits but with border-line significance. The resulting best-fits, 
however, all concur in indicating a
flatter component emerging at higher energies, with spectral changes $\Delta
\alpha_{\rm x} \sim 0.8$ around $1-3$ keV. 

\item the comparison between \sax and ROSAT spectra (Fig. 4). Excluding 4C
56.27, which appears to have an extremely flat ROSAT spectrum, all our LBL
have $\alpha_{\rm x}(ROSAT) > \alpha_{\rm x}(BeppoSAX)$, with a typical
difference $\sim 0.7$. Although the interpretation of this difference is
complicated by possible spectral variability effects, it is unlikely that
these can explain the fact that five out of six of our LBL have a ROSAT spectrum 
which is
steeper than the \sax one. Similarly, possible ROSAT miscalibrations, if at
all present, could only explain a difference $\sim 0.2 - 0.3$.

\item the spectral energy distributions (Fig. 5 and 6). Despite the
non-simultaneity of the multifrequency data (UMRAO radio data excluded) it is
apparent that the \sax spectra indicate a different emission component in the
SEDs of our LBL sources, separate from that responsible for the low energy
emission. In fact, the extrapolation of the relatively flat \sax slopes cannot
be extended to much lower frequencies since the predicted optical flux would
be orders of magnitude below the observed value. A sharp steepening towards
lower frequencies is then necessary to meet the much higher optical
(synchrotron) flux, as also required by the comparison with ROSAT data.
Detailed synchrotron inverse Compton model fits to the SEDs fully confirm this
picture and constrain the physical parameters in these sources (Tab. 7).

\item the $\alpha_{\rm x} - \nu_{\rm peak}$ diagram (Fig. 7). Our
interpretation of this plot is the one originally proposed by Padovani \&
Giommi (1996) for the ROSAT data. Namely, $\alpha_{\rm x}$ is flat for LBL due
to the dominance of the inverse Compton emission, and steepens moving from LBL
to HBL as synchrotron replaces inverse Compton as the main emission mechanism
in the X-ray band. The spectral index then flattens again as the synchrotron
peak moves to higher energies in the X-ray band, eventually converging to the
relatively flat value characteristic of synchrotron emission before the
peak. Again, this fits perfectly with a dominance of inverse Compton emission
in our LBL.

\end{itemize}

Note that strong, direct evidence for the presence of both synchrotron and
inverse Compton emission in the \sax~spectra of LBL has been presented by
Giommi et al. (1999) and Tagliaferri et al. (2000) for S5 0716+714 and ON 231,
respectively. All data for the two HBL included in this study are consistent
with synchrotron emission extending into the \sax band for these sources, in
agreement with the results of Wolter et al. (1998).

\sax data for four more 1 Jy BL Lacs have been obtained and data reduction is
in progress. Those results will be presented in a forthcoming paper, where we
will address the properties of the full 1Jy \sax sample and their physical
parameters in more details.

\section*{Acknowledgements}
We thank F. Fiore for providing the code to de-redden the XSPEC unfolded
spectra used to construct the spectral energy distribution plots and Tommaso
Maccacaro, Franco Mantovani, and Carlo Stanghellini for their contribution at
an early stage of this project. AC acknowledges financial support from
ASI-ARS-99-75 while LC acknowledges the STScI Visitor Program. This research
has made use of data from the University of Michigan Radio Astronomy
Observatory which is supported by funds from the University of Michigan and of
the NASA/IPAC Extragalactic Database (NED), which is operated by the Jet
Propulsion Laboratory, California Institute of Technology, under contract with
the National Aeronautics and Space Administration.


\begin{thebibliography}{}
\bibitem{} Boella G. \etal, 1997a, A\&AS, 122, 299
\bibitem{} Boella G. \etal, 1997b, A\&AS, 122, 327
\bibitem{} Comastri A., Molendi S., Ghisellini G., 1995, MNRAS, 277, 297 
\bibitem{} Comastri A., Fossati G., Ghisellini G., Molendi S., 1997, ApJ, 
480, 534  
\bibitem{} Dickey J. M., Lockman F. J., 1990, ARAA, 28, 215
\bibitem{} Elvis M., Lockman F. J., Wilkes, B. J., 1989, AJ, 97, 777
\bibitem{} Fossati G., Maraschi L., Celotti A., Comastri A., Ghisellini G., 1998,
MNRAS, 299, 433  
\bibitem{} Fossati G. et al., 2000, ApJ, 541, 166 
\bibitem{} Frontera F. et al., 1997, A\&AS, 122, 357
\bibitem{} Gehrels N., 1986, ApJ, 303, 336
\bibitem{} Ghisellini G., Celotti A., Fossati G., Maraschi L., Comastri A.,
1998, MNRAS, 301, 451        
\bibitem{} Giommi P., Fiore, F., 1997, in 5th International Workshop on
Data Analysis in Astronomy, Erice, in press 
\bibitem{} Giommi P. et al., 1998, A\&A, 333, L5
\bibitem{} Giommi P. et al., 1999, A\&A, 351, 59
\bibitem{} Giommi P., Padovani P., 1994, MNRAS, 268, L51
\bibitem{} Giommi P., Padovani P., Perlman E., 2000, MNRAS, 317, 743 
\bibitem{} Heidt J., Wagner S. J., 1996, A\&A, 305, 42 
\bibitem{} Iwasawa K., Fabian A. C., Nandra K., 1999, MNRAS, 307, 611
\bibitem{} Kollgaard R. I., 1994, Vistas Astron., 38, 29 
\bibitem{} Kubo H., Takahashi T., Madejski G., Tashiro M., Makino F., 
Inoue S., Takahara F., 1998, ApJ, 504, 693  
\bibitem{} Lamer G., Brunner H., Staubert R., 1996, A\&A, 311, 384 
\bibitem{} Lin Y. C. et al., 1999, ApJ, 525, 191 
\bibitem{} Lucas R., Liszt H. S., 1993, A\&A, 276, L33 
\bibitem{} Mineo T. et al., 2000, A\&A, 359, 471 
\bibitem{} Morrison R., McCammon D., 1983, ApJ, 270, 119
\bibitem{} Padovani P., Giommi P., 1995, ApJ, 444, 567  
\bibitem{} Padovani P., Giommi P., 1996, MNRAS, 279, 526  
\bibitem{} Parmar A. et al., 1997, A\&AS, 122, 309.
\bibitem{} Perlman E. S., Madejski G., Stocke J. T., Rector T. A., 1999,
ApJ, 523, L11 
\bibitem{} Pian E. et al., 1998, ApJ, 492, L17 
\bibitem{} Sambruna R., Maraschi L., Urry C. M., 1996, ApJ, 463, 444
\bibitem{} Sambruna R., Ghisellini G., Hooper E., Kollgaard R. I., Pesce 
J. E., Urry C. M., 1999, ApJ, 515, 140 
\bibitem{} Spada M., Lazzati D., Ghisellini G., Celotti A., 2001, MNRAS,
in press (astro-ph/0103424) 
\bibitem{} Stickel M., Padovani P., Urry C. M., Fried J. W., K\"uhr H., 1991,
ApJ, 374, 431 
\bibitem{} Tagliaferri G. et al., 2000, A\&A, 354, 431  
\bibitem{} Tagliaferri G. et al., 2001, A\&A, 368, 38 
\bibitem{} Tanihata C., et al., 2000, ApJ, 543, 124 
\bibitem{} Tapia S., Craine E. R., Johnson K., 1976, ApJ, 203, 291  
\bibitem{} Tavecchio F., Maraschi L., Ghisellini G., 1998, ApJ, 509, 608  
\bibitem{} Urry C. M., Padovani, P., 1995, PASP, 107, 803 
\bibitem{} Urry C. M., Sambruna R. M., Worrall D. M., Kollgaard R. I., 
Feigelson E. D., Perlman E. S., Stocke J. T., 1996, ApJ, 463, 424 
\bibitem{} White R. L., Giommi P., Angelini L., 1995, 
{\tt http://lheawww.gsfc.nasa.gov/users/white/wgacat/wgacat.html}
\bibitem{} Wolter A. et al., 1998, A\&A, 335, 899 

\end{thebibliography}
\end{document}